\documentclass[sigconf]{acmart}
\AtBeginDocument{%
  }

\acmISBN{978-1-4503-XXXX-X/18/06}

\usepackage{algorithm}
\usepackage{algpseudocode}
\usepackage{multirow}
\usepackage{booktabs}
\usepackage{array} 
\begin{document}

\title{Unlocking Real-Time Fluorescence Lifetime Imaging: Multi-Pixel Parallelism for FPGA-Accelerated Processing}

\author{Ismail Erbas}
\authornote{Both authors contributed equally to this research.}
\email{erbasi@rpi.edu}
\affiliation{%
  \institution{Center for Modeling, Simulation, \& Imaging in Medicine}
  \institution{Rensselaer Polytechnic Institute}
  \city{Troy}
  \state{NY}
  \country{USA}
}

\author{Aporva Amarnath}
\authornotemark[1]
\email{aporva.amarnath@ibm.com}
\affiliation{%
  \institution{IBM T.J. Watson Research Center}
  \city{Yorktown Heights}
  \state{NY}
  \country{USA}
}

\author{Vikas Pandey}
\email{pandev2@rpi.edu}
\affiliation{%
  \institution{Center for Modeling, Simulation, \& Imaging in Medicine}
  \institution{Rensselaer Polytechnic Institute}
  \city{Troy}
  \state{NY}
  \country{USA}
}
\author{Karthik Swaminathan}
\email{kvswamin@us.ibm.com}
\affiliation{%
  \institution{IBM T.J. Watson Research Center}
  \city{Yorktown Heights}
  \state{NY}
  \country{USA}
}
\author{Naigang Wang}
\email{nwang@us.ibm.com}
\affiliation{%
  \institution{IBM T.J. Watson Research Center}
  \city{Yorktown Heights}
  \state{NY}
  \country{USA}
}

\author{Xavier Intes}
\email{intesx@rpi.edu}
\affiliation{%
  \institution{Center for Modeling, Simulation, \& Imaging in Medicine}
  \institution{Rensselaer Polytechnic Institute}
  \city{Troy}
  \state{NY}
  \country{USA}
}


\begin{abstract}
Fluorescence lifetime imaging (FLI) is a widely used technique in the biomedical field for measuring the decay times of fluorescent molecules, providing insights into metabolic states, protein interactions, and ligand-receptor bindings. However, its broader application in fast biological processes, such as dynamic activity monitoring, and clinical use, such as in guided surgery, is limited by long data acquisition times and computationally demanding data processing. While deep learning has reduced post-processing times, time-resolved data acquisition remains a bottleneck for real-time applications. To address this, we propose a method to achieve real-time FLI using an FPGA-based hardware accelerator. Specifically, we implemented a GRU-based sequence-to-sequence (Seq2Seq) model on an FPGA board compatible with time-resolved cameras. The GRU model balances accurate processing with the resource constraints of FPGAs, which have limited DSP units and BRAM. The limited memory and computational resources on the FPGA require efficient scheduling of operations and memory allocation to deploy deep learning models for low-latency applications. We address these challenges by using STOMP, a queue-based discrete-event simulator that automates and optimizes task scheduling and memory management on hardware. By integrating a GRU-based Seq2Seq model and its compressed version, called Seq2SeqLite, generated through knowledge distillation, we were able to process multiple pixels in parallel, reducing latency compared to sequential processing. We explore various levels of parallelism to achieve an optimal balance between performance and resource utilization. Our results indicate that the proposed techniques achieved a 17.7x and 52.0x speedup over manual scheduling for the Seq2Seq model and the Seq2SeqLite model, respectively.
\end{abstract}




\maketitle

\section{Introduction}
In the biomedical field, non-invasive imaging techniques that effectively capture dynamic cellular processes are crucial for advancing disease detection, monitoring treatment responses, and developing new therapies. Fluorescence lifetime imaging (FLI) has become a valuable tool in this context, providing detailed insights into cellular and molecular activities by measuring the decay times of fluorescent signals \cite{verma2024fluorescence}. Unlike traditional intensity-based fluorescence imaging, FLI remains unaffected by variables such as fluorophore concentration and excitation intensity. This makes FLI highly suitable for investigating complex biological phenomena, including protein-protein interactions and ligand-receptor binding in intact organisms, especially in cancer imaging \cite{dmitriev2021luminescence,yuan2024antibody,verma2024using}. This work aims to bridge the gap between the computational demands of advanced FLI data analysis and the practical limitations of existing hardware platforms used for deployment. By leveraging efficient scheduling techniques and optimizing resource allocation, we strive to make real-time FLI more accessible for clinical and research applications. This work thus lays the foundation for implementing deep learning models on hardware-constrained devices for biomedical applications, moving us closer to the goal of real-time computational imaging solutions that can keep pace with rapid biological processes. 

FLI requires a time-resolved detector, such as an Intensified Charge-Coupled Device (ICCD) \cite{yuan2024experimental}, Time-Correlated Single-Photon Counting (TCSPC), or a time-gated Single-Photon Avalanche Diode (SPAD) \cite{bruschini2019single}. A fluorescent sample is excited with a short pulse of light, and the emitted photons are captured over time, generating a \emph{Temporal Point Spread Function (TPSF)}, which represents the distribution of photon arrival times after the excitation pulse. Fluorescence decay follows first-order kinetics, so the sample decay can be modeled as the summation of the exponential decays of all the fluorophore components present in the sample, called \emph{Sample Decay Function (SDF)}. However, the observed decay function, or TPSF, is the convolution of the SDF and the \emph{Instrument Response Function (IRF)}, which represents the temporal response of the imaging system to an instantaneous input \cite{chen2019vitro}.
Accurately extracting the SDF and its decay parameters from the observed TPSF remains computationally challenging. Traditional methods, such as nonlinear least-squares fitting, maximum likelihood estimation, and center-of-mass calculations, are computationally intensive and often rely on initial parameter guesses \cite{becker2012fluorescence}. These approaches are iterative and require prior estimation of the IRF before re-convolution and fitting can take place, making them less practical for applications that demand rapid FLI parameter estimation.
Recent developments in deep learning methods offer promising alternatives to these conventional techniques \cite{smith2019fast, erbas2024fluorescence,Nizam2024}, providing direct estimation of the decay parameters. In particular, Recurrent Neural Networks (RNNs), especially GRU-based models \cite{pandey2024deep}, have been adapted to demonstrate fast deconvolution in order to estimate the temporal SDF and its components from the TPSF without requiring prior estimation of the IRF. The Seq2Seq model is adept at handling noisy sequential time-series data and can significantly reduce computational complexity compared to traditional methods, making real-time analysis more feasible.
The implementation of this GRU-based deep learning model into field-programmable gate arrays (FPGAs) introduces new challenges. The complexities of the models and the computational demands can strain the limited availability of memory and processing resources in FPGAs. This issue becomes more pronounced when dealing with large-format time-resolved SPAD detectors, which feature a high-resolution $512 \times 512$ pixel array \cite{bruschini2019single}. The substantial data volume generated by such detectors increases the demands on the FPGA's resources, complicating real-time processing efforts.
These challenges can be addressed through efficient scheduling of computation operations and efficient management of memory allocations. 
For this purpose, we adapted STOMP~\cite{vega2020stomp}, a tool used for discrete event simulations in multicore processors, for optimizing scheduling and data processing in resource-constrained FPGAs.
Through optimal scheduling of tasks and managing resource allocation, we were able to organize computations in order to maximize parallel processing capabilities on the camera-compatible FPGA, while adhering to its hardware constraints.
By simulating various scheduling policies for the operations and memory processes required for the GRU-based Seq2Seq model, we designed the most efficient way to process multiple pixels in parallel, significantly reducing latency. This approach thus allowed us to maximize the number of pixels that can be processed simultaneously without exceeding the FPGA's resource limitations, thus enabling us to achieve near-real time processing capability. This paper, thus, makes the following contributions: \\
\noindent $\bullet$ We propose a GRU-based Seq2Seq model specifically optimized for real-time FLI on constrained FPGAs\\
\noindent $\bullet$ We introduce a scheduling mechanism for efficiently processing tasks on resource-constrained, camera-compatible FPGAs, leveraging pixel-level parallelism \\
\noindent $\bullet$ We incorporate KD and QAT to create a lightweight, hardware-friendly model (Seq2SeqLite) that significantly reduces computational complexity\\
\noindent $\bullet$ Our approach achieves near real-time processing, enabling FLI estimation in under 500 ms, suitable for dynamic biological processes and clinical implementation.

\section{FLI time-series estimation}
\begin{figure}[t]
    \centering
    \includegraphics[width=\columnwidth]{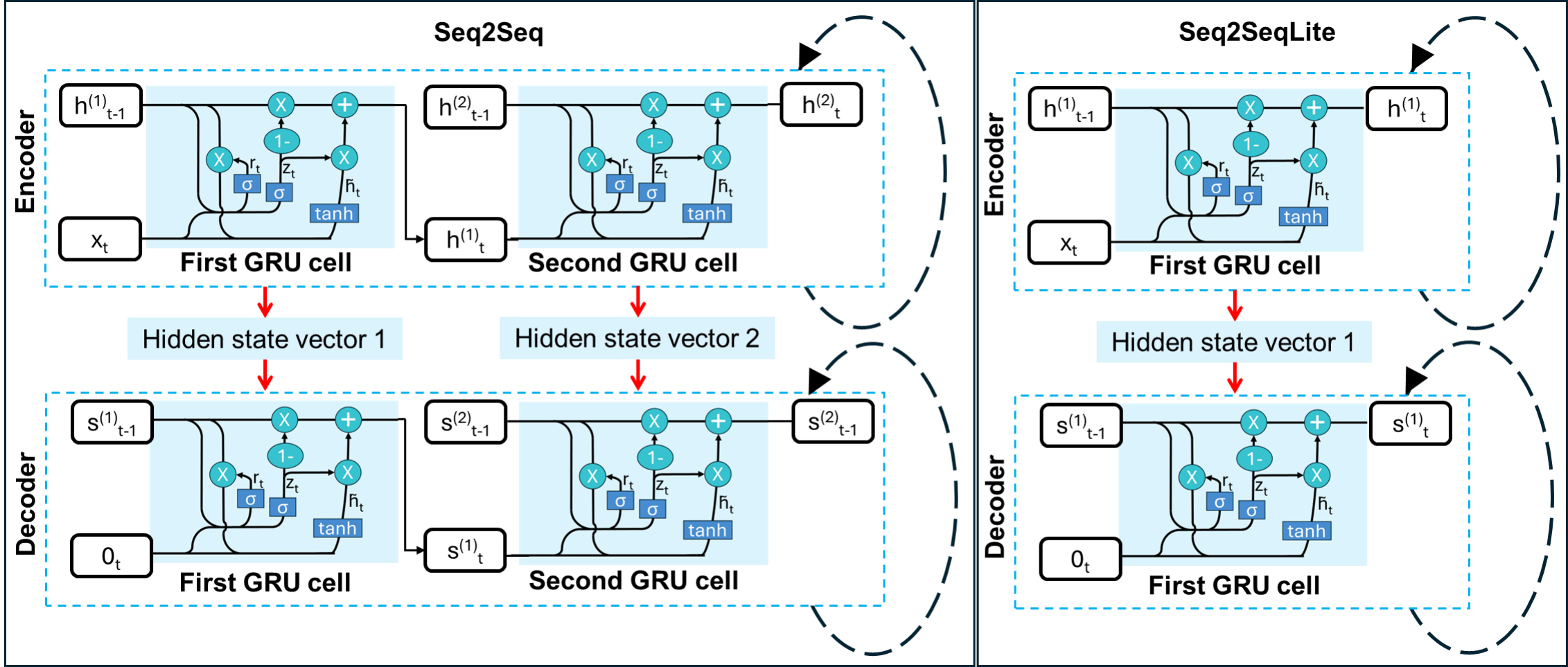} 
    \caption{ (Left) Deep Sequence-to-Sequence (Seq2Seq) model featuring two GRU cells in both the encoder and decoder, with 128 hidden units in the first GRU cell of the encoder. The encoder processes the input time sequence and passes its final hidden states to initialize the decoder, which updates its hidden states to generate the output. (Right) Seq2SeqLite model, a compressed version of the deep Seq2Seq model, generated using quantization-aware training (QAT) combined with knowledge distillation (KD).}
    \vspace{-0.2in}
    \label{fig:seq2seq_arch}
\end{figure}

We implemented the GRU-based sequence-to-sequence (Seq2Seq) architecture from \cite{pandey2024deep} for fast estimation of SDF sequences from TPSF in FLI (Fig.~\ref{fig:seq2seq_arch}). This model processes the TPSF time-series input for each pixel and outputs the corresponding SDF sequence, enabling efficient real-time analysis without explicit deconvolution or prior knowledge of the IRF. This deep Seq2Seq model consists of an encoder and a decoder layers, each comprising two GRU cells with 128 hidden units. The initial hidden states of all GRU cells are set to zero vectors of size 128, for the encoder, while in the decoder, hidden states are initialized based on the final hidden states of the encoder GRU cells.

In the encoder, the first GRU cell processes the input TPSF sequence $\{\mathbf{x}_1, \mathbf{x}_2, \dots, \mathbf{x}_T\}$, where $T$ is the number of time gates (70 in this case). At each time step $t$, it updates its hidden state $\mathbf{h}_t^{(1)}$ using the previous hidden state $\mathbf{h}_{t-1}^{(1)}$ and the input $\mathbf{x}_t$:

\begin{align}
    \mathbf{z}_t^{(1)} &= \sigma\left( \mathbf{W}_z^{(1)} \mathbf{x}_t + \mathbf{U}_z^{(1)} \mathbf{h}_{t-1}^{(1)} + \mathbf{b}_z^{(1)} \right), \\
    \mathbf{r}_t^{(1)} &= \sigma\left( \mathbf{W}_r^{(1)} \mathbf{x}_t + \mathbf{U}_r^{(1)} \mathbf{h}_{t-1}^{(1)} + \mathbf{b}_r^{(1)} \right), \\
    \tilde{\mathbf{h}}_t^{(1)} &= \tanh\left( \mathbf{W}_h^{(1)} \mathbf{x}_t + \mathbf{U}_h^{(1)} \left( \mathbf{r}_t^{(1)} \odot \mathbf{h}_{t-1}^{(1)} \right) + \mathbf{b}_h^{(1)} \right), \\
    \mathbf{h}_t^{(1)} &= \left( 1 - \mathbf{z}_t^{(1)} \right) \odot \mathbf{h}_{t-1}^{(1)} + \mathbf{z}_t^{(1)} \odot \tilde{\mathbf{h}}_t^{(1)},
    \label{first_gru_enc}
\end{align}

where:
\begin{itemize}
    \item $\mathbf{z}_t^{(1)}$ is the update gate vector.
    \item $\mathbf{r}_t^{(1)}$ is the reset gate vector.
    \item $\tilde{\mathbf{h}}_t^{(1)}$ is the candidate hidden state.
    \item $\sigma(\cdot)$ denotes the sigmoid activation function.
    \item $\tanh(\cdot)$ denotes the hyperbolic tangent activation function.
    \item $\odot$ denotes element-wise multiplication.
    \item $\mathbf{W}_*^{(1)}$, $\mathbf{U}_*^{(1)}$, and $\mathbf{b}_*^{(1)}$ are learned weight matrices and bias vectors for the first GRU cell in the encoder.
\end{itemize}

The second GRU cell in the encoder takes output from the first GRU cell $\mathbf{h}_t^{(1)}$ as its input and updates its own hidden state $\mathbf{h}_t^{(2)}$:

\begin{align}
    \mathbf{z}_t^{(2)} &= \sigma\left( \mathbf{W}_z^{(2)} \mathbf{h}_t^{(1)} + \mathbf{U}_z^{(2)} \mathbf{h}_{t-1}^{(2)} + \mathbf{b}_z^{(2)} \right), \\
    \mathbf{r}_t^{(2)} &= \sigma\left( \mathbf{W}_r^{(2)} \mathbf{h}_t^{(1)} + \mathbf{U}_r^{(2)} \mathbf{h}_{t-1}^{(2)} + \mathbf{b}_r^{(2)} \right), \\
    \tilde{\mathbf{h}}_t^{(2)} &= \tanh\left( \mathbf{W}_h^{(2)} \mathbf{h}_t^{(1)} + \mathbf{U}_h^{(2)} \left( \mathbf{r}_t^{(2)} \odot \mathbf{h}_{t-1}^{(2)} \right) + \mathbf{b}_h^{(2)} \right), \\
    \mathbf{h}_t^{(2)} &= \left( 1 - \mathbf{z}_t^{(2)} \right) \odot \mathbf{h}_{t-1}^{(2)} + \mathbf{z}_t^{(2)} \odot \tilde{\mathbf{h}}_t^{(2)},
    \label{second_gru_enc}
\end{align}

After processing all time steps, we obtain the final hidden states $\mathbf{h}_T^{(1)}$ and $\mathbf{h}_T^{(2)}$ from the encoder's two GRU cells. These hidden states are passed to the decoder as initial states for its GRU cells.

In the decoder, the first GRU cell uses the final hidden state $\mathbf{h}_T^{(1)}$ from the encoder's first GRU cell as its initial hidden state. The input to this first GRU cell is a zero vector at each time step:

\begin{align}
    \mathbf{z}_t^{(1)'} &= \sigma\left( \mathbf{W}_z^{(1)'} \mathbf{0} + \mathbf{U}_z^{(1)'} \mathbf{s}_{t-1}^{(1)} + \mathbf{b}_z^{(1)'} \right), \\
    \mathbf{r}_t^{(1)'} &= \sigma\left( \mathbf{W}_r^{(1)'} \mathbf{0} + \mathbf{U}_r^{(1)'} \mathbf{s}_{t-1}^{(1)} + \mathbf{b}_r^{(1)'} \right), \\
    \tilde{\mathbf{s}}_t^{(1)} &= \tanh\left( \mathbf{W}_h^{(1)'} \mathbf{0} + \mathbf{U}_h^{(1)'} \left( \mathbf{r}_t^{(1)'} \odot \mathbf{s}_{t-1}^{(1)} \right) + \mathbf{b}_h^{(1)'} \right), \\
    \mathbf{s}_t^{(1)} &= \left( 1 - \mathbf{z}_t^{(1)'} \right) \odot \mathbf{s}_{t-1}^{(1)} + \mathbf{z}_t^{(1)'} \odot \tilde{\mathbf{s}}_t^{(1)},
    \label{first_gru_dec}
\end{align}

where $\mathbf{s}_0^{(1)} = \mathbf{h}_T^{(1)}$. Here, $\mathbf{W}_*^{(1)'}$, $\mathbf{U}_*^{(1)'}$, and $\mathbf{b}_*^{(1)'}$ are the updated weights and biases for the first GRU cell in the decoder.

The second GRU cell in the decoder takes the hidden state from the first GRU cell $\mathbf{s}_t^{(1)}$ as its input, and its initial hidden state is set to the final hidden state $\mathbf{h}_T^{(2)}$ from the encoder:

\begin{align}
    \mathbf{z}_t^{(2)'} &= \sigma\left( \mathbf{W}_z^{(2)'} \mathbf{s}_t^{(1)} + \mathbf{U}_z^{(2)'} \mathbf{s}_{t-1}^{(2)} + \mathbf{b}_z^{(2)'} \right), \\
    \mathbf{r}_t^{(2)'} &= \sigma\left( \mathbf{W}_r^{(2)'} \mathbf{s}_t^{(1)} + \mathbf{U}_r^{(2)'} \mathbf{s}_{t-1}^{(2)} + \mathbf{b}_r^{(2)'} \right), \\
    \tilde{\mathbf{s}}_t^{(2)} &= \tanh\left( \mathbf{W}_h^{(2)'} \mathbf{s}_t^{(1)} + \mathbf{U}_h^{(2)'} \left( \mathbf{r}_t^{(2)'} \odot \mathbf{s}_{t-1}^{(2)} \right) + \mathbf{b}_h^{(2)'} \right), \\
    \mathbf{s}_t^{(2)} &= \left( 1 - \mathbf{z}_t^{(2)'} \right) \odot \mathbf{s}_{t-1}^{(2)} + \mathbf{z}_t^{(2)'} \odot \tilde{\mathbf{s}}_t^{(2)},
    \label{second_gru_dec}
\end{align}

\noindent where $\mathbf{s}_0^{(2)}$ = $\mathbf{h}_T^{(2)}$. The output SDF sequence is generated by passing the hidden state $\mathbf{s}_t^{(2)}$ through a fully connected (dense) layer:

\begin{equation}
    \mathbf{y}_t = \mathbf{W}_o \mathbf{s}_t^{(2)} + \mathbf{b}_o,
\end{equation}

where $\mathbf{W}_o$ and $\mathbf{b}_o$ are the learned weight matrix and bias vector of the dense layer. The computation is summarized in the Algorithm~\ref{alg:algorithm1}.

\begin{algorithm}
\caption{Seq2Seq GRU Model for SDF Estimation}\label{alg:algorithm1}
\begin{algorithmic}[1]
\State \textbf{Input:} TPSF sequence $\{\mathbf{x}_1, \mathbf{x}_2, \dots, \mathbf{x}_T\}$
\State \textbf{Output:} SDF sequence $\{\mathbf{y}_1, \mathbf{y}_2, \dots, \mathbf{y}_T\}$

\State Initialize encoder hidden states: $\mathbf{h}_0^{(1)} \leftarrow \mathbf{0}$, $\mathbf{h}_0^{(2)} \leftarrow \mathbf{0}$

\For{$t \gets 1$ to $T$}
    \State $\mathbf{h}_t^{(1)} \leftarrow \text{GRU}^{(1)}_{\text{enc}}(\mathbf{x}_t, \mathbf{h}_{t-1}^{(1)})$
    \State $\mathbf{h}_t^{(2)} \leftarrow \text{GRU}^{(2)}_{\text{enc}}(\mathbf{h}_t^{(1)}, \mathbf{h}_{t-1}^{(2)})$
\EndFor

\State Initialize decoder hidden states: $\mathbf{s}_0^{(1)} \leftarrow \mathbf{h}_T^{(1)}$, $\mathbf{s}_0^{(2)} \leftarrow \mathbf{h}_T^{(2)}$

\For{$t \gets 1$ to $T$}
    \State $\mathbf{s}_t^{(1)} \leftarrow \text{GRU}^{(1)}_{\text{dec}}(\mathbf{0}, \mathbf{s}_{t-1}^{(1)})$
    \State $\mathbf{s}_t^{(2)} \leftarrow \text{GRU}^{(2)}_{\text{dec}}(\mathbf{s}_t^{(1)}, \mathbf{s}_{t-1}^{(2)})$
    \State $\mathbf{y}_t \leftarrow \mathbf{W}_o \mathbf{s}_t^{(2)} + \mathbf{b}_o$
\EndFor
\end{algorithmic}
\end{algorithm}

\paragraph{Seq2SeqLite Model:}
In this study, we utilized \emph{Knowledge Distillation (KD)} and \emph{Quantization-Aware Training} \cite{erbas2024compressingrecurrentneuralnetworks} to generate a compressed and simplifed version of Seq2Seq architectures. Through this process, a smaller Seq2SeqLite model Fig.~\ref{fig:seq2seq_arch} with a single GRU layer in both the encoder and decoder was trained to mimic the performance of a larger teacher model. The Seq2SeqLite model offers reduced computational complexity, making it ideal for deployment in hardware-constrained environments such as FPGAs. In line with findings of \cite{erbas2024compressingrecurrentneuralnetworks}, the 32x32 Seq2SeqLite model was identified as the most efficient configuration, balancing both accuracy and hardware resource consumption. Consequently, this specific version was chosen for deployment in our FPGA implementation.

\paragraph{Fluorescence Lifetime Calculation:}
Once the SDF and its exponential decay components are estimated, the decay parameters can be calculated using simple calculations. For each pixel in the spatial domain, the fluorescence lifetime (\(\tau\)) is computed using the trapezoidal rule.  In the case of a bi-exponential model, two fluorescence lifetime values (\(\tau_1\) and \(\tau_2\)) can be determined using trapezoidal numerical integration. 
The integration of the time-resolved signal across the gates yields the lifetime value. For each pixel \((i, j)\), the lifetime value \(\tau_1(i, j)\) is computed as follows:

\begin{equation}
\tau_{1}(i,j) = \frac{1}{\text{S}_{\text{max}}(i,j)} \sum_{k=2}^{\text{gate}} \left( t(k) - t(k-1) \right) \frac{\text{S}(i,j,k) + \text{S}(i,j,k-1)}{2}
\end{equation}

Here, \(\text{S}_{\text{max}}(i,j)\) represents the maximum or initial signal at the pixel \((i,j)\), and the trapezoidal integration is performed over the time vector \(t\) and the corresponding signal values. The same procedure is applied to compute \(\tau_2(i, j)\) for the second set of time-resolved data.

\section{Description}
In this section, we detail our FPGA implementation and our parallel execution method utilizing a scheduler adapted for FPGAs.

\subsection{FPGA Implementation}
For efficient memory utilization, the available BRAM was divided into three distinct memory types: constant memory, shared memory, and data memory. 

\paragraph{Constant Memory} Constant Memory stores the model weights. For the Seq2Seq model, the weights consist of $\mathbf{W} \in \mathbb{R}^{1 \times 128}$ for the first GRU cell and $\mathbf{W} \in \mathbb{R}^{128 \times 128}$ for the second GRU cell. The matrices $\mathbf{U} \in \mathbb{R}^{128 \times 128}$ represent the recurrent weights for all GRU cells, and the biases are represented by vectors $\mathbf{b} \in \mathbb{R}^{1 \times 128}$. In the Seq2SeqLite model, the weights are reduced: $\mathbf{W} \in \mathbb{R}^{1 \times 32}$ for both the encoder and decoder GRU layers, $\mathbf{U} \in \mathbb{R}^{32 \times 32}$ for all cells, and $\mathbf{b} \in \mathbb{R}^{1 \times 32}$.

\paragraph{Shared Memory} Shared Memory stores the hidden state vectors of the GRU cells. For the Seq2Seq model, the hidden states are $\mathbf{h} \in \mathbb{R}^{1 \times 128}$ for both the encoder and decoder's first and second GRU cells. In the Seq2SeqLite model, the hidden state vectors are reduced to $\mathbf{h} \in \mathbb{R}^{1 \times 32}$ for both encoder and decoder GRU cells.

\paragraph{Data Memory} Data Memory is used for temporary storage of intermediate results generated during the computations within the GRU cells. This memory facilitates the calculation of the hidden layers by caching intermediate operations, such as the candidate hidden states and gate activations during the GRU computations.

The segmentation of memory resources, as described, allows for efficient parallel processing and reduces memory access contention, thereby optimizing the overall system performance on the FPGA.\autoref{tab:fpga_resources} summarizes the resource utilization for both the Seq2Seq and Seq2SeqLite models, including memory requirements and DSP utilization.

\paragraph{DSP Operations on FPGA} The FPGA's DSP blocks are responsible for performing arithmetic operations such as addition, multiplication, division, and activation functions (e.g., sigmoid, tanh). The operations within the GRU cells are computationally intensive, particularly in the matrix-vector multiplications for the weight and recurrent weight matrices $\mathbf{W}$ and $\mathbf{U}$ (see \autoref{tab:fpga_resources} for details on DSP usage). For example, the Seq2Seq model requires 171,080 multiply operations and 172,340 add operations per inference cycle. The DSP blocks handle these operations in parallel, thus significantly speeding up computations. In the Seq2SeqLite model, the DSP usage is reduced due to the smaller matrix sizes, allowing for a lower computational load and more efficient utilization of the FPGA resources. This combined strategy of memory segmentation and efficient DSP utilization allows us to meet the real-time processing requirements for GRU-based models, enabling high-performance operation on FPGA hardware.

\begin{table}[h!]
\centering
\setlength{\tabcolsep}{5pt} 
\renewcommand{\arraystretch}{1.1} 
\scriptsize
\begin{tabular}{@{}llcc@{}}
\toprule
\textbf{FPGA Components} & \textbf{Operation/Data} & \textbf{Seq2Seq Model} & \textbf{Seq2SeqLite Model} \\
\midrule
\multirow{6}{*}{\textbf{DSP}} & Add & 172340 & 17150 \\
 & Multiply & 171080 & 16310 \\
 & Divide & 1050 & 630 \\
 & Subtract & 283 & 143 \\
 & Sigmoid & 560 & 280 \\
 & Tanh & 280 & 140 \\
\midrule
\textbf{Constant Memory (BRAM)} & Model weights& $128 \times 128$ ($400$ KB)& $32 \times 32$ ($6.73$ KB)\\
\textbf{Data Memory (BRAM)} & Intermediate data& $128 \times 128$  & $32 \times 32$ \\
\textbf{Shared Memory (BRAM)} & Hidden states& $128 \times 128$  ($0.25$ KB)& $32 \times 32$ ($0.06$ KB)\\
\bottomrule
\end{tabular}
\caption{FPGA resource usage for Seq2Seq and Seq2SeqLite models.}
\label{tab:fpga_resources}
 \vspace{-0.3in}
\end{table}

\subsection{Exploiting Pixel-Level Parallelism through Scheduling}

\begin{figure}[t]
    \centering
    \includegraphics[width=\columnwidth]{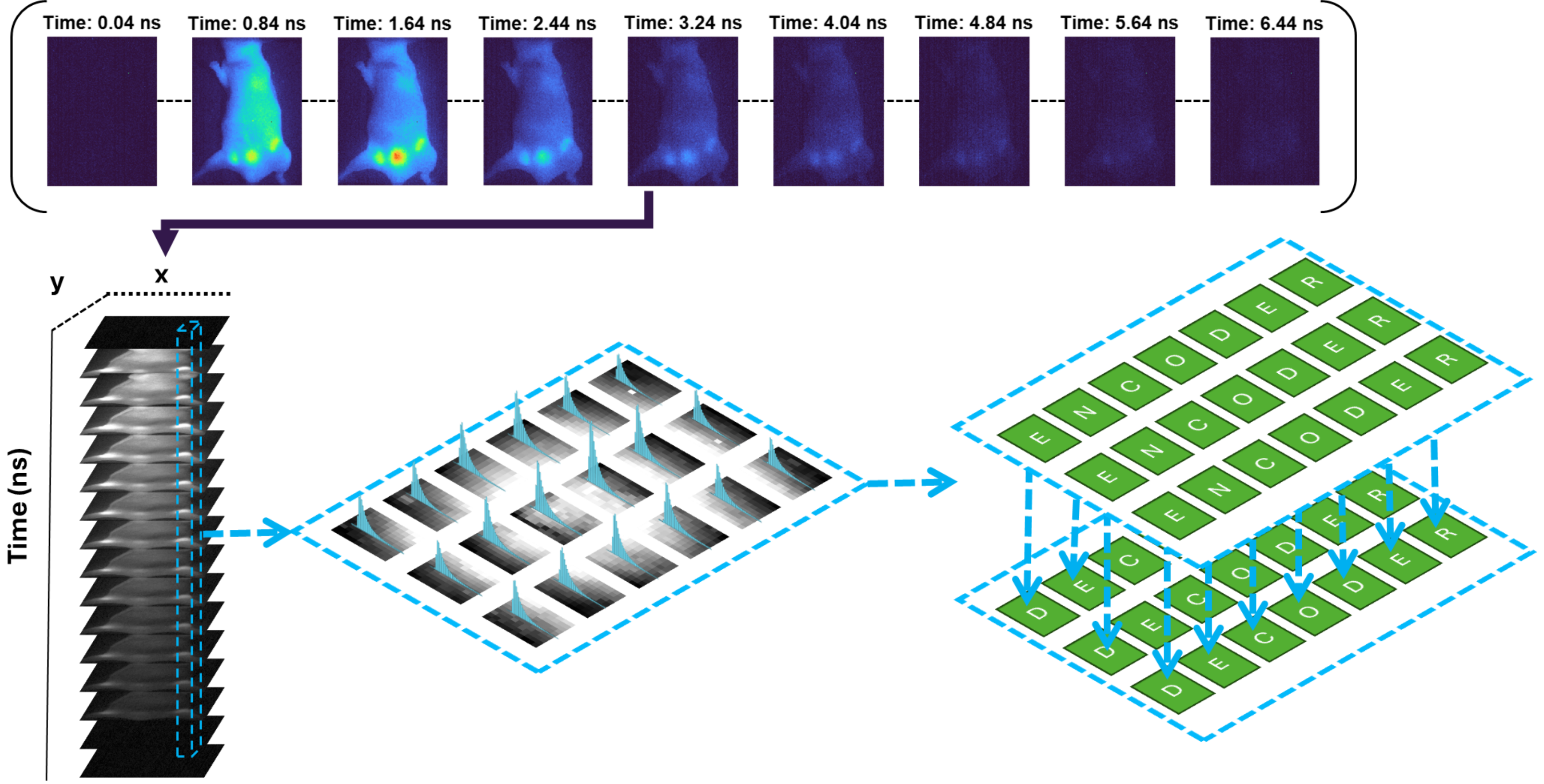} 
    \caption{(Top) Time-series data points captured using a large-format time-resolved camera in Macroscopic Fluorescence Lifetime Imaging (MFLI). These experimental, wide-field, time-resolved images illustrate the entire body of a nude mouse containing HER2+ tumor xenografts labeled with Alexa Fluor 700 conjugated to Trastuzumab for a non-invasive drug-target binding MFLI study. (Bottom) Schematic representation of parallel data processing for multiple pixels with different temporal point spread functions (TPSFs)}
    \label{fig:figure_parallel}
     \vspace{-0.2in}
\end{figure}

Time-series data, unlike traditional image processing, presents opposing data flow patterns. Each image taken per time step is to be processed sequentially and cannot be batched and executed in parallel. However, time-series calculation allows for pixel-level parallelism. Fig.~\ref{fig:figure_parallel} showcases this. On the top, we see an image is captured every 40~ns. To compute this time series data, we have to compute each particular pixel across images captured in time sequentially (the blue tower in the bottom left). Each time step of this pixel represents a temporal point in its TPSF, which is sequentially processed through the encoder and decoder GRU cells. This process is inherently sequential because different regions of the TPSF contain information about the various fluorophores present in the sample. However, across the x and y dimensions, we can process the data in parallel to exploit spatial independence. We exploit this parallelism to improve our FLI calculation time while significantly improving FPGA utilization.

\begin{figure}[t]
    \centering
    \includegraphics[width=0.8\columnwidth]{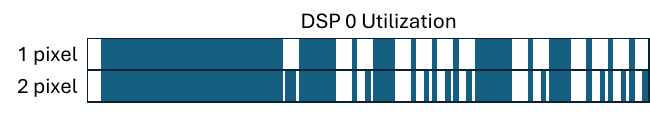} 
    \caption{Utilization of DSP 0 of 128 when executing the encoder layer on one pixel vs. two pixels in parallel.}
    \label{fig:gantt_seq_vs_parallel}
     \vspace{-0.2in}
\end{figure}

To further demonstrate our motivation for exploiting pixel-level parallelism, we compare the utilization of one of the DSPs on the FPGA when we execute one pixel at a time \textit{vs.} two pixels in parallel. As shown in Fig.~\ref{fig:gantt_seq_vs_parallel} (top), when one pixel's data is computed through the encoder it utilizes 61.5\% of DSP 0 and 1.8\% of all the DSPs (not shown). As the number of pixels are processed in parallel we can improve the utilization of the DSPs and the BRAMs. An example of this can be seen in Fig.~\ref{fig:gantt_seq_vs_parallel} (bottom) for the encoder simultaneously operating on 2 pixels, achieving a utilization of 69.6\% for DSP 0 and 3.5\% for all DSPs (not shown). We can see that some of the underutilized slots on the DSP are now utilized to compute for the second pixel.

For this, our methodology utilizes a modified static scheduling and mapping policy that is built to improve utilization of the FPGA for better performance. We adapt STOMP~\cite{vega2020stomp}, a tool for discrete event simulations in multicore processors, to optimize scheduling and data processing in resource-constrained FPGAs. For this, we first generate graphs for the encoder, decoder and post processing parts of the application. Using the simulator, we discover ready tasks for each graph. The ready tasks are then mapped and scheduled onto the DSPs and BRAMs using a modified non-blocking first-come first-serve scheduling policy. For the mapping of the ready tasks, we assign a ready task to the next available resource of a kind on the FPGA (DSPs and BRAMs) that the task can execute on. The operations on the DSP are detailed in Table~\ref{tab:fpga_resources}. Further, the application include reads and writes from/to the BRAMs utilized for constant, data and shared memories.

We perform this mapping and scheduling for multiple pixels being computed simultaneously by providing a trace of the application graph per pixel. We then choose the number of pixels that can be operated on in parallel and its schedule based on the configuration maximizes resource utilization and performance. 

\section{Methodology}  
\paragraph{Knowledge Distillation and Seq2SeqLite Model Training} We followed the same training process outlined in \cite{erbas2024compressingrecurrentneuralnetworks}, the Seq2Seq model acts as the teacher, and the Seq2SeqLite model serves as the student. The student model is trained to minimize a combined loss function, incorporating both standard task loss and distillation loss, to ensure it learns the behaviors of the larger model while retaining efficiency. Furthermore, QAT was employed to simulate 8-bit precision during training. This allows the Seq2SeqLite model to adapt to reduced precision computations, ensuring minimal accuracy degradation after quantization. QAT prepares the model for the resource constraints of FPGA deployment, where low-precision arithmetic significantly reduces power consumption and memory usage. The model was trained using the Adam optimizer with a learning rate of 0.001 and Mean Squared Error (MSE) as a loss function. The model was implemented using Tensorflow~\cite{tensorflow} and trained on a system with an Intel i9-11900KF CPU and a Nvidia RTX 3090 GPU.

\paragraph{FPGA Specifications for Deployment} The XEM7360-K410T FPGA, based on the Xilinx Kintex-7 XC7K410T, was selected due to its compatibility with the time-resolved SPAD array camera and its ability to meet the specific computational and memory requirements of the models. The FPGA includes 1540 DSP slices, which handle the arithmetic operations necessary for matrix multiplications and other calculations in the GRU layers. These slices allow for the efficient execution of the mathematical operations required for recurrent layers during inference. Additionally, the FPGA has 795 Block RAM (BRAM) tiles, used to store intermediate data such as activations, weights, and hidden states. Given the sequential nature of the GRU-based model, where hidden states need to be passed across time steps, the BRAM tiles were crucial for managing memory during the execution of the model, allowing for efficient data flow between the encoder and decoder. These specifications provided the necessary resources for deploying the quantized Seq2SeqLite model, which was optimized to 8-bit precision. The DSP slices managed the computational load, while the BRAM handled the memory demands of the time-series data. 

\paragraph{Scheduler Specifications} 
STOMP is a queue-based discrete event simulator that allows for plug-and-play scheduling policies~\cite{stomp2021}. While it was built to optimize scheduling for heterogeneous multiprocessors and accelerators, we adapt it to perform scheduling on FPGAs. The tasks of an application are expressed as operations listed in Table~\ref{tab:fpga_resources} on the DSPs and read and write from/to BRAMs. We created data flow graphs for a pixel for the encoder, decoder and post processing. To execute and schedule parallel pixels, we provide a trace of graphs arriving at the hardware simultaneously.

\section{Empirical Evaluation}
In this section, we evaluate the effectiveness of our scheduling methodology for parallel pixel processing on FPGAs, comparing it to a manually scheduled single pixel execution. We assess both the Seq2Seq model and its compressed and quantized variant, Seq2SeqLite, to demonstrate improvements in terms of execution time and hardware utilization. For our evaluation, we utilized 128 DSPs, 128 BRAMs for the constant memory, 128 BRAMs for the shared memory and 256 BRAMs for the data memory.

\begin{figure}[t!]
    \centering
    \includegraphics[width=\linewidth]{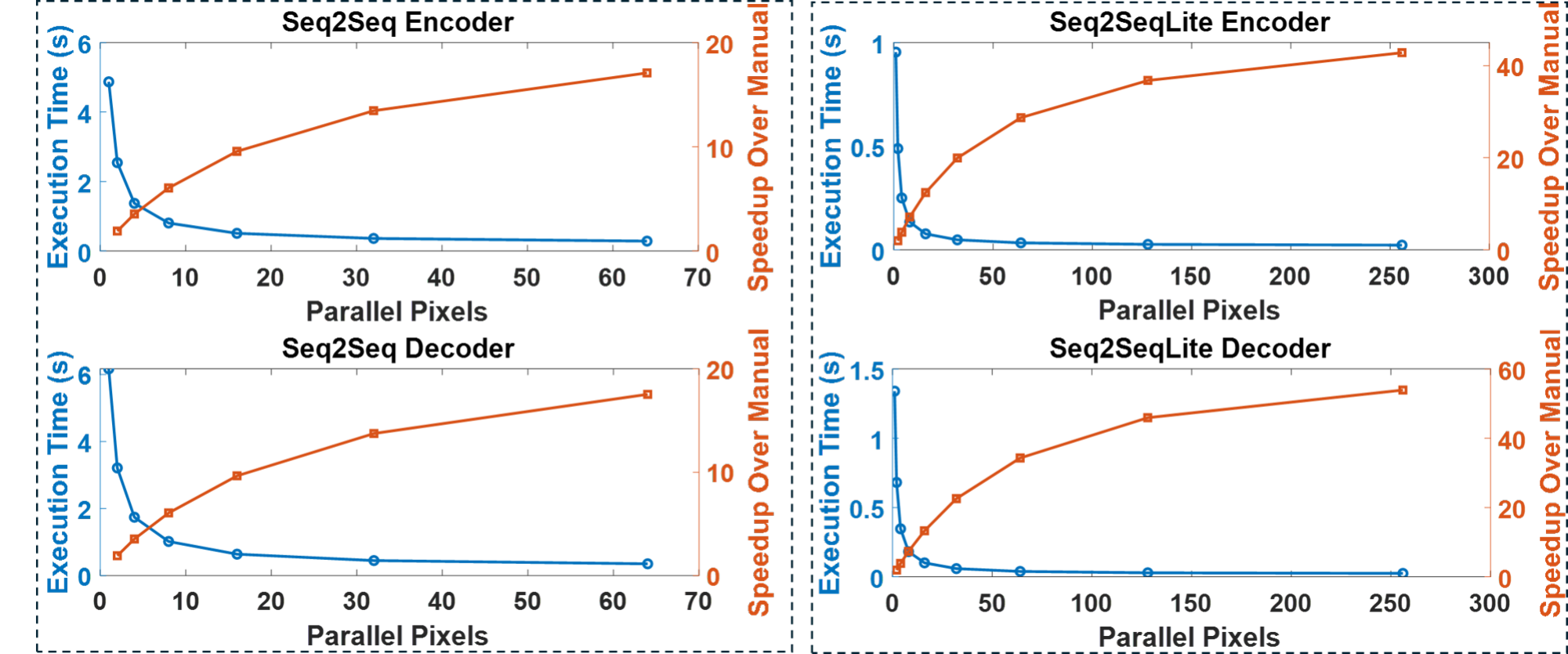} 
    \caption{Parallel execution results for Seq2Seq (left) and Seq2SeqLite (right) models. The top row shows the number of pixels processed in parallel for the encoder, and the bottom row illustrates the same for the decoder. Both execution time and speedup over manual processing are displayed for different parallel pixel configurations.}
    \label{fig:combined_results}
     \vspace{-0.1in}
\end{figure}

\begin{figure}[t!]
    \includegraphics[width=1.01\columnwidth]{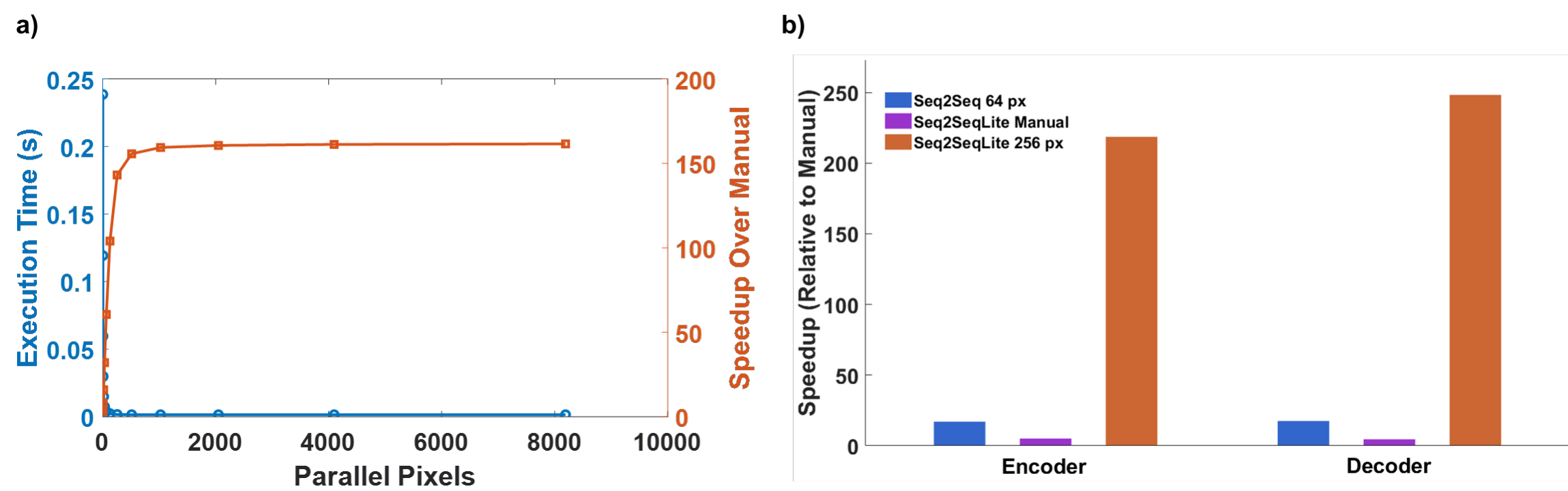} 
    \caption{(a) Parallel execution results for the post processing step. Both execution time and speedup over manual processing are displayed for different parallel pixel configurations (b) Speedup comparison between Seq2Seq and Seq2SeqLite models. The bars represent the relative speedup for three configurations: Seq2Seq with 64 pixels, Seq2SeqLite (manual), and Seq2SeqLite with 256 pixels. Results are shown separately for the encoder and decoder stages.}
    \label{fig:pp_barplot}
     \vspace{-0.1in}
\end{figure}

As shown in Fig.~\ref{fig:combined_results}, the proposed parallel execution strategy significantly accelerates the processing of FLI tasks. For the Seq2Seq model, we are able to process 64 pixels in parallel for both the encoder and decoder stages. This allows us to achieve a maximum speedup of 17.1$\times$ and 17.5$\times$ over the manual schedule for the encoder and decoder layers, respectively. The Seq2SeqLite model showed even greater improvements. We are able to execute 256 pixels in parallel and achieve speedups of 42.8$\times$ and 53.9$\times$, respectively. Furthermore, when we schedule pixels in parallel for the post processing step (see Fig.~\ref{fig:pp_barplot}(a)), we are able to achieve a speedup 160.6$\times$ for 512 parallel pixels over single pixel execution. Also note that processing any more pixels in parallel doesn't provide more speedup as the hardware utilization is saturated.

As shown in Fig.~\ref{fig:pp_barplot}(b), our parallel pixel execution approach yields a speedup of 17.7$\times$ for the Seq2Seq model, while the reduced and quantized Seq2SeqLite model alone provides a speedup of 4.5$\times$. However, neither of these methods individually is sufficient to achieve real-time FLI execution. When combined, the efficiency of Seq2SeqLite with the parallel pixel execution strategy results in a total speedup of 230$\times$, enabling FLI estimation in under 500 ms, thus meeting the requirements for near real-time performance. 

\begin{figure}[b]
    \centering
    \includegraphics[width=1\columnwidth]{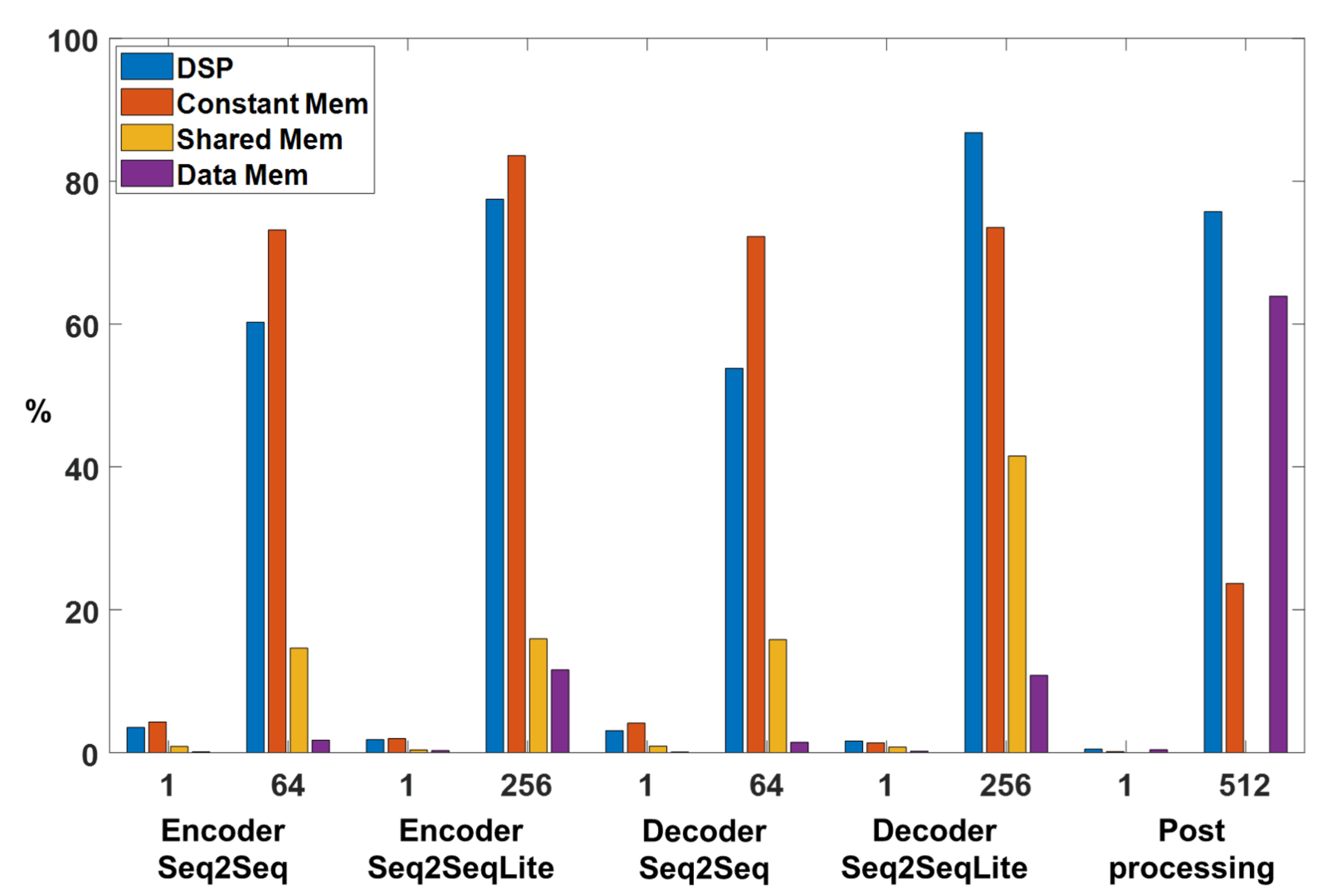} 
    \caption{Utilization comparison between sequential manual scheduling and parallel pixel scheduled execution for encoder, decoder and post processing for Seq2Seq and Seq2SeqLite models.}
    \label{fig:utilization}
\end{figure}

Our analysis shown in Fig.~\ref{fig:utilization} demonstrates that by utilizing our parallel pixel execution method, we are able to significantly increase the hardware utilization. We are able to achieve a DSP utilization of 57.0\% and 80.0\% for the Seq2Seq and Seq2SeqLite models, respectively. We also observe that constant memory is heavily utilized in the encoder and decoder execution, while data memory is better utilized during the post processing phase. 

Based on these observations, as future work, we propose to further improve our performance by designing LUT-based execution on the FPGA for certain operations listed in Table.~\ref{tab:fpga_resources}, thereby increasing our compute resources, and allowing us to process even more pixels in parallel. We also plan to adopt a different distribution of BRAMs used for constant, shared and data memory during the different phases of encoder, decoder and post processing.

\section{Related Work}
Numerous approaches\cite{erbas2024,demirkiran2022ensemble,erbas2024compressingrecurrentneuralnetworks,pandey2024deep,smith2019fast,bengio1994learning,chung2014empirical,vaswani2017attention,liu2024lightweight,wang2020linformer,cho2014learning,lin2024coupling,balti2024bi,das2023recurrent,erbacs2024real} have been proposed for time-series estimation across different fields. Earlier methods focused on statistical models like ARIMA \cite{jain2017study}, followed by the adoption of RNNs, such as LSTMs and GRUs, due to their ability to capture long-term dependencies in sequential data \cite{bengio1994learning,chung2014empirical}. Recently, transformer-based architectures have gained popularity for time-series applications. With their self-attention mechanism, transformers excel in capturing relationships over long sequences \cite{vaswani2017attention}. However, they require a large number of parameters, often in the range of millions, making them less suitable for resource-constrained environments, such as FPGAs or edge devices \cite{liu2024lightweight,wang2020linformer}. 

Given the challenges of deploying large transformer models in low-resource settings, GRUs have been used as a more practical alternative. GRUs offer a balance between model complexity and performance by reducing the number of parameters compared to LSTMs and transformers while still providing robust time-series estimation \cite{cho2014learning}. For instance, GRU-based encoder-decoder architectures have been applied to handle time-series data in various domains, showing their suitability for FPGA implementations \cite{erbas2024compressingrecurrentneuralnetworks,balti2024bi,lin2024coupling,das2023recurrent}. This makes GRU models particularly advantageous for applications requiring efficient use of limited computational and memory resources, such as real-time FLI.  In a recent study \cite{lin2024coupling}, researchers designed a compact FLI system based on Piccolo SPAD sensors and implemented a GRU-based recurrent model on an FPGA platform for real-time lifetime parameter estimation. This approach achieved significant reductions in latency and data bandwidth, making real-time FLI feasible. Similarly, another study \cite{erbas2024compressingrecurrentneuralnetworks} focuses on compressing GRU neural networks for real-time FLI on FPGAs. To address the resource constraints of FPGAs, the authors employ compression techniques such as post-training quantization, quantization-aware training, and KD. These techniques result in a smaller model, called Seq2SeqLite, which reduces the model size significantly while maintaining accuracy. The compressed model is tested on experimental data, achieving real-time performance with 8-bit precision, demonstrating that GRU-based models can improve the performance of FLI for clinical and research applications. However, none of these prior work explore parallel pixel execution approach on time-series data while utilizing an optimized scheduling policy to perform FLI estimation on a resource-constrained FPGA to achieve real-time execution.

\section{Conclusion}
In this work, we have demonstrated the feasibility of achieving real-time FLI on resource-constrained FPGAs by leveraging efficient scheduling techniques and pixel-level parallelism. Our approach significantly improves DSP and BRAM utilization, allowing multiple pixels to be processed in parallel, thereby reducing latency and enabling real-time performance. By integrating a GRU-based Seq2Seq model and its compressed counterpart, Seq2SeqLite, we optimized the FPGA’s limited resources, achieving significant speedups through automated scheduling on parallel pixels.

Our results show a substantial improvement in the execution time of both the encoder and decoder stages, achieving 17.7$\times$ and 52.0$\times$ speedup for real-time FLI estimation for a 299k and 7k parameter models, respectively. The combination of deep learning model compression, quantization-aware training, and parallel pixel execution has proven highly effective for this application. This research provides a critical step toward making real-time FLI more accessible for clinical and biomedical applications, such as guided surgery and dynamic biological process monitoring.

Future work will explore further optimizations in FPGA resource management, such as LUT-based execution, and alternative memory distribution strategies to enhance performance and allow for processing even more pixels in parallel. These enhancements will continue to push the boundaries of real-time computational imaging on constrained hardware platforms, ultimately achieving video-level performance, making it suitable for clinical implementation.


\bibliographystyle{plain}
\bibliography{sample-base}

\begin{thebibliography}{10}

\bibitem{tensorflow}
Tensorflow.
\newblock \url{tensorflow.org}.

\bibitem{balti2024bi}
Hanen Balti, Ali Ben~Abbes, and Imed~Riadh Farah.
\newblock A bi-gru-based encoder--decoder framework for multivariate time
  series forecasting.
\newblock {\em Soft Computing}, pages 1--12, 2024.

\bibitem{becker2012fluorescence}
Wolfgang Becker.
\newblock Fluorescence lifetime imaging--techniques and applications.
\newblock {\em Journal of microscopy}, 247(2):119--136, 2012.

\bibitem{bengio1994learning}
Yoshua Bengio, Patrice Simard, and Paolo Frasconi.
\newblock Learning long-term dependencies with gradient descent is difficult.
\newblock {\em IEEE transactions on neural networks}, 5(2):157--166, 1994.

\bibitem{bruschini2019single}
Claudio Bruschini, Harald Homulle, Ivan~Michel Antolovic, Samuel Burri, and
  Edoardo Charbon.
\newblock Single-photon avalanche diode imagers in biophotonics: review and
  outlook.
\newblock {\em Light: Science \& Applications}, 8(1):87, 2019.

\bibitem{chen2019vitro}
Sez-Jade Chen, Nattawut Sinsuebphon, Alena Rudkouskaya, Margarida Barroso,
  Xavier Intes, and Xavier Michalet.
\newblock In vitro and in vivo phasor analysis of stoichiometry and
  pharmacokinetics using short-lifetime near-infrared dyes and time-gated
  imaging.
\newblock {\em Journal of biophotonics}, 12(3):e201800185, 2019.

\bibitem{cho2014learning}
Kyunghyun Cho.
\newblock Learning phrase representations using rnn encoder-decoder for
  statistical machine translation.
\newblock {\em arXiv preprint arXiv:1406.1078}, 2014.

\bibitem{chung2014empirical}
Junyoung Chung, Caglar Gulcehre, KyungHyun Cho, and Yoshua Bengio.
\newblock Empirical evaluation of gated recurrent neural networks on sequence
  modeling.
\newblock {\em arXiv preprint arXiv:1412.3555}, 2014.

\bibitem{das2023recurrent}
Susmita Das, Amara Tariq, Thiago Santos, Sai~Sandeep Kantareddy, and Imon
  Banerjee.
\newblock Recurrent neural networks (rnns): architectures, training tricks, and
  introduction to influential research.
\newblock {\em Machine Learning for Brain Disorders}, pages 117--138, 2023.

\bibitem{demirkiran2022ensemble}
Ferhat Demirk{\i}ran, Aykut {\c{C}}ay{\i}r, U{\u{g}}ur {\"U}nal, and Hasan
  Da{\u{g}}.
\newblock An ensemble of pre-trained transformer models for imbalanced
  multiclass malware classification.
\newblock {\em Computers \& Security}, 121:102846, 2022.

\bibitem{dmitriev2021luminescence}
Ruslan~I Dmitriev, Xavier Intes, and Margarida~M Barroso.
\newblock Luminescence lifetime imaging of three-dimensional biological
  objects.
\newblock {\em Journal of Cell Science}, 134(9):1--17, 2021.

\bibitem{erbacs2024real}
{\.I}smail Erba{\c{s}} and Burak G{\"u}{\c{c}}l{\"u}.
\newblock Real-time vibrotactile pattern generation and identification using
  discrete event-driven feedback.
\newblock {\em Somatosensory \& Motor Research}, 41(2):77--89, 2024.

\bibitem{erbas2024compressingrecurrentneuralnetworks}
Ismail Erbas, Vikas Pandey, Aporva Amarnath, Naigang Wang, Karthik Swaminathan,
  Stefan~T. Radev, and Xavier Intes.
\newblock Compressing recurrent neural networks for fpga-accelerated
  implementation in fluorescence lifetime imaging.
\newblock 2024.

\bibitem{erbas2024fluorescence}
Ismail Erbas, Vikas Pandey, Navid~Ibtehaj Nizam, Nanxue Yuan, and Xavier Intes.
\newblock Fluorescence lifetime parameters estimation with transformer based
  deep learning model.
\newblock In {\em Clinical and Translational Biophotonics}, pages TS3B--2.
  Optica Publishing Group, 2024.

\bibitem{erbas2024}
Ismail Erbas, Vikas Pandey, Navid~Ibtehaj Nizam, Nanxue Yuan, Amit Verma,
  Margarida Barroso, and Xavier Intes.
\newblock Transformer-based deep learning model for fluorescence lifetime
  parameter estimations using pixelwise instrument response function.
\newblock {\em Research Square}, October 2024.

\bibitem{stomp2021}
IBM.
\newblock Stomp: Scheduling techniques optimization in heterogeneous
  multi-processors.
\newblock \url{https://github.com/ibm/stomp}, 2021.
\newblock Accessed: 2024-09-03.

\bibitem{jain2017study}
Garima Jain and Bhawna Mallick.
\newblock A study of time series models arima and ets.
\newblock {\em Available at SSRN 2898968}, 2017.

\bibitem{lin2024coupling}
Yang Lin, Paul Mos, Andrei Ardelean, Claudio Bruschini, and Edoardo Charbon.
\newblock Coupling a recurrent neural network to spad tcspc systems for
  real-time fluorescence lifetime imaging.
\newblock {\em Scientific Reports}, 14(1):3286, 2024.

\bibitem{liu2024lightweight}
Hou-I Liu, Marco Galindo, Hongxia Xie, Lai-Kuan Wong, Hong-Han Shuai, Yung-Hui
  Li, and Wen-Huang Cheng.
\newblock Lightweight deep learning for resource-constrained environments: A
  survey.
\newblock {\em ACM Computing Surveys}, 2024.

\bibitem{Nizam2024}
Navid~Ibtehaj Nizam, Vikas Pandey, Ismail Erbas, Jason~T. Smith, and Xavier
  Intes.
\newblock A novel technique for fluorescence lifetime tomography.
\newblock {\em bioRxiv}, 2024.

\bibitem{pandey2024deep}
Vikas Pandey, Ismail Erbas, Xavier Michalet, Arin Ulku, Claudio Bruschini,
  Edoardo Charbon, Margarida Barroso, and Xavier Intes.
\newblock Deep learning-based temporal deconvolution for photon time-of-flight
  distribution retrieval.
\newblock 2024.

\bibitem{smith2019fast}
Jason~T Smith, Ruoyang Yao, Nattawut Sinsuebphon, Alena Rudkouskaya, Nathan Un,
  Joseph Mazurkiewicz, Margarida Barroso, Pingkun Yan, and Xavier Intes.
\newblock Fast fit-free analysis of fluorescence lifetime imaging via deep
  learning.
\newblock {\em Proceedings of the national academy of sciences},
  116(48):24019--24030, 2019.

\bibitem{vaswani2017attention}
A~Vaswani.
\newblock Attention is all you need.
\newblock {\em Advances in Neural Information Processing Systems}, 2017.

\bibitem{vega2020stomp}
Augusto Vega, Aporva Amarnath, John-David Wellman, Hiwot Kassa, Subhankar Pal,
  Hubertus Franke, Alper Buyuktosunoglu, Ronald Dreslinski, and Pradip Bose.
\newblock Stomp: A tool for evaluation of scheduling policies in heterogeneous
  multi-processors.
\newblock {\em arXiv preprint arXiv:2007.14371}, 2020.

\bibitem{verma2024fluorescence}
Amit Verma, Vikas Pandey, Catherine Sherry, Christopher James, Kailie Matteson,
  Jason~T Smith, Alena Rudkouskaya, Xavier Intes, and Margarida Barroso.
\newblock Fluorescence lifetime imaging for quantification of targeted drug
  delivery in varying tumor microenvironments.
\newblock {\em bioRxiv}.

\bibitem{verma2024using}
Amit Verma, Catherine Sherry, Nanxue Yuan, Vikas Pandey, John Williams, Xavier
  Intes, and Margarida~M Barroso.
\newblock Using meditope-based antibody labeling to improve fluorescence
  lifetime fret imaging.
\newblock In {\em Multiphoton Microscopy in the Biomedical Sciences XXIV}, page
  PC128470S. SPIE, 2024.

\bibitem{wang2020linformer}
Sinong Wang, Belinda~Z Li, Madian Khabsa, Han Fang, and Hao Ma.
\newblock Linformer: Self-attention with linear complexity.
\newblock {\em arXiv preprint arXiv:2006.04768}, 2020.

\bibitem{yuan2024experimental}
Nanxue Yuan, Vikas Pandey, Xavier Michalet, and Xavier Intes.
\newblock Experimental study of fluorescence lifetime uncertainty in time-gated
  iccd-based macroscopic fluorescence lifetime imaging.
\newblock In {\em Clinical and Translational Biophotonics}, pages TM5B--4.
  Optica Publishing Group, 2024.

\bibitem{yuan2024antibody}
Nanxue Yuan, Vikas Pandey, Amit Verma, John~C Williams, Xavier Intes, and
  Margarida Barroso.
\newblock Antibody-target binding quantification in living tumors using
  macroscopy fluorescence lifetime forster resonance energy transfer imaging
  (mfli fret).
\newblock In {\em Visualizing and Quantifying Drug Distribution in Tissue
  VIII}, volume 12821, pages 17--20. SPIE, 2024.

\end{thebibliography}

\end{document}